# Signature of BKT-like spin transport in a quasi-2D antiferromagnet $BaNi_2V_2O_8$

Kurea Nakagawa[1], Minoru Kanega[2], Tomoyuki Yokouchi[1], Masahiro Sato[2], and Yuki Shiomi[1]

[1]*Department of Basic Science, The University of Tokyo, Tokyo 153-8902, Japan*
[2]*Department of Physics, Chiba University, Chiba, 263-8522, Japan*

*Abstract*

In two-dimensional (2D) spin systems, the augmentation of spin fluctuations gives rise to quasi-long-range order; however, how they manifest in spin transport remains unclear. Here we investigate the spin Seebeck effect (SSE) in a quasi-2D antiferromagnet, $BaNi_2V_2O_8$, which has been reported to exhibit the Berezinskii-Kosterlitz-Thouless (BKT) transition owing to its distinct 2D nature. We found that the SSE in $Pt/BaNi_2V_2O_8$ persists well above the Néel temperature, significantly different from the behavior of 3D ordered magnets. Our numerical analysis for a 2D microscopic spin model supports the hypothesis that the observed SSE is linked to strong magnetic correlations in the BKT-like phase.

*Introduction-* Spin current generation [1] is a crucial technique finding its applications in spintronics and next-generation information processing. Not only does it have engineering significance, but it is also fundamentally important as an effective method for elucidating the spin excitation and transport of magnetic states. One of the most versatile methods for driving spin current is the spin Seebeck effect (SSE) [2–5]. The SSE refers to the generation of a spin current in a magnetic material in response to a temperature gradient across the junction between a magnetic material and a metal. Thanks to its simple bilayer structure and straightforward nature, the SSE serves as a valuable tool for investigating the spin dynamics in diverse magnetic insulators, including ferromagnets [6,7], ferrimagnets [8–10], paramagnets [11,12], antiferromagnets [13–18], and noncollinear spin systems [19,20]. More recently, the SSE has also been applied to exotic magnetic systems, such as a 1D quantum spin liquid [21], a spin-nematic liquid [22], a spin-Peierls magnet [23], and a magnon-BEC magnet [24], revealing unique spin transport in low-dimensional spin systems.

In the realm of exotic magnetic dynamics, two-dimensional (2D) spin systems are particularly intriguing because of their enhanced spin fluctuations. In the case of an ideal 2D Heisenberg magnet, long-range magnetic order is forbidden at finite temperatures [25,26]; instead, it exhibits a quasi-long-range order in a magnetic field, accompanied by the Berezinskii-Kosterlitz-Thouless (BKT) transition [27–31]. The BKT transition is known to exhibit very weak singularities (Namely, it may be viewed as an invisible transition). While BKT transitions and their related topics in 2D or quasi-2D magnets have long been explored mainly focusing on thermodynamic quantities, the effect of spin fluctuations in such systems on transport and nonequilibrium properties remains elusive. Therefore, it is intriguing to experimentally study spin transport in 2D systems by using a spin current that is sensitive to spin dynamics. Although SSE in a layered ferromagnetic insulator with weak anisotropy in exchange coupling strength has been reported [32], highly 2D systems exhibiting BKT-like behavior have not yet been experimentally explored. Furthermore, since very recent theoretical works have revealed a novel contribution of the BKT spin texture to the spin current [33,34], the experimental study of spin current transport in BKT magnets is timely.

As a promising candidate for realizing two-dimensional (2D) magnetic systems with BKT transitions, $BaNi_2V_2O_8$ stands out as a rare example in which a BKT-like transition has been experimentally confirmed. $BaNi_2V_2O_8$ is a quasi-2D Heisenberg antiferromagnet ($T_N$ = 47.75 ± 0.25 K) with weak XY anisotropy [35,36]. Magnetic susceptibility and specific heat do not show any sharp transitions associated with long-range magnetic ordering [35,36]. This pronounced 2D magnetic characteristic is mainly due to its outstandingly weak interlayer interaction ($|J_{out}| < 10^{-4}J_1$, where $J_{out}$ and $J_1$ are interlayer exchange interaction and nearest-neighbor intralayer exchange interaction, respectively) [36]. Its trigonal crystal structure houses spin-1 $Ni^{2+}$ magnetic ions, forming honeycomb layers stacked parallel to the $c$-axis [Fig. 1(a)]. As illustrated in Fig. 1(b), spins lie within the honeycomb plane in the ground state, displaying a dominant antiferromagnetic exchange interaction between the nearest-neighbor $Ni^{2+}$ ions. The BKT-like transition has been experimentally confirmed using electron spin resonance [37], nuclear magnetic resonance [38], and neutron scattering measurements [39], reporting that the BKT-like transition temperature ($T_{BKT}$) lies between 40 and 45 K, just below $T_N$.

In this study, we have investigated SSE in $BaNi_2V_2O_8$: a quasi-2D antiferromagnet with a BKT-like transition. The SSE of Pt/$BaNi_2V_2O_8$-(100) significantly persists even above the Néel temperature, without any anomalies at the transition temperature, in contrast to the behavior of three-dimensional (3D) ordered magnets. Interestingly, these results are consistent with the calculation results for a weakly-anisotropic 2D Heisenberg antiferromagnet with the BKT transition, which does not show long-range order. Therefore, the SSE in this system is plausibly linked to the strong magnetic correlations in the BKT-like phase over a very broad temperature range, despite $BaNi_2V_2O_8$ exhibiting magnetic order at low temperatures in a static sense.

*Methods*- Single crystals of $BaNi_2V_2O_8$ were grown by a flux method (see also Supplemental Material (SM) [41]). The resultant crystals with yellow hexagonal plates were confirmed the $BaNi_2V_2O_8$ stoichiometry by x-ray fluorescence measurements (Hitachi, EA6000VX). Magnetization was measured using a superconducting magnet Quantum Design MPMS3.

The SSE devices used in the present study consist of a 5 nm thick Pt film sputtered on top of a $BaNi_2V_2O_8$ single crystal. The samples have a surface size of 0.9 × 0.25-0.4 mm$^2$ with a thickness of 0.23-0.29 mm. In this study, the SSE was measured using a self-heating method [40] in a Quantum Design PPMS9. The Pt layer was utilized not only as a detection layer for spin currents, but also as a heater to generate thermal gradients. An a.c. current $I_c \propto \sin \omega t$ was applied to the Pt film and generates a temperature gradient across Pt/$BaNi_2V_2O_8$ bilayer due to the Joule heating. Thermally induced spin current with $2\omega$ frequency produces inverse spin Hall effect signal in the Pt layer as a second harmonic voltage. By measuring the second-harmonic voltage $V_{2\omega}$ using a lock-in amplifier (NF LI5650) under magnetic fields applied perpendicular to both a.c. current and temperature gradient, we can selectively detect the SSE signal. The typical parameters of the a.c. current were the RMS amplitude of 4.5-6 μA and the frequency of 83 Hz (see also SM [41]).

*Results*-We first examined the magnetic properties of $BaNi_2V_2O_8$. As shown in Fig. 1(c), magnetization shows a linear magnetic-field dependence over the entire temperature range. Temperature dependence of magnetic susceptibility $\chi$, shown in Fig. 1 (d), is indicative of 2D spin properties; it shows a characteristic behavior with a very broad peak around $T_{max} \sim 143$ K. The magnitude of $\chi$ is almost the same for $H \parallel a$ and $H \parallel c$ at high temperatures, suggesting isotropic magnetism. At temperatures below the broad maximum, the susceptibility gradually decreases, and the anisotropy becomes more pronounced as temperature decreases. A small tail observed at very low temperatures is attributed to paramagnetic impurities, which have been frequently observed in this material [35,36] and other low-dimensional materials [42,43]. The Néel temperature is determined to be $T_N = 47$ K from the minimum point of $\frac{d \chi_{H \parallel c}}{d T}$ [Fig. 1(e)], following previous studies [36,39]. It is notable that the large $T_{max} / T_N$ ratio highlights low-dimensional magnetism, and strong short-range correlations remain far above the transition temperature. The overall magnetic properties are consistent with those reported in previous studies [35,36,39]. The BKT transition temperature is invisible in magnetic susceptibility.

The SSE for $BaNi_2V_2O_8$ was then measured for two different configurations, as illustrated in

Figs. 2(a) and (b): Pt films were sputtered on the (100)- or (001)- crystalline plane of $BaNi_2V_2O_8$, which corresponds to the configuration where the 2D honeycomb plane is perpendicular or parallel to the Pt plane, respectively. In Fig. 2 (c), the magnetic field dependence of $V_{2\omega}$ in Pt/$BaNi_2V_2O_8$-(100) and Pt/$BaNi_2V_2O_8$-(001) at 20 K ($< T_N$) is shown along with that in a Pt/$SiO_2$ control sample. Here $SiO_2$ means a thermally oxidized Si substrate that is diamagnetic. Therefore, it is expected that Pt/$SiO_2$ will exhibit a voltage solely originating from the normal Nernst effect of Pt [10]. As shown in Fig. 2(c), the observed $V_{2\omega}$ of Pt/$SiO_2$ is almost independent of the magnetic field, showing that the Nernst effect in Pt is too small to be detected in our devices. In contrast to Pt/$SiO_2$, both Pt/$BaNi_2V_2O_8$ devices show positive monotonic magnetic-field dependences, which can be attributed to the SSE signal. The positive sign of the SSE signals is the same as that of Pt/YIG, a typical ferrimagnetic SSE material (see SM [41]). We further verified that W (10 nm) / $BaNi_2V_2O_8$-(100) shows the opposite sign of voltage response at 10 K, consistent with the opposite sign of the spin Hall angle in W [41]. Notably, the signal of Pt/$BaNi_2V_2O_8$-(100) is larger than that of Pt/$BaNi_2V_2O_8$-(001). This fact is in line with the 2D magnetic character of $BaNi_2V_2O_8$: For the Pt/$BaNi_2V_2O_8$-(100) configuration, spin current generated within the 2D plane can directly flow into the Pt, while for the Pt/$BaNi_2V_2O_8$-(001) configuration, the spin current hardly flows due to the very weak interplane interaction.

Next, we show in Fig. 2 (d) the results of the same measurement conducted at 100 K, well above the Néel temperature. At this temperature, Pt/$SiO_2$ and Pt/$BaNi_2V_2O_8$-(001) show almost no magnetic-field dependence within the confidence range. Nevertheless, Pt/$BaNi_2V_2O_8$-(100) still shows a finite SSE signal with a linear positive dependence on the magnetic field. At this temperature, long-range magnetic order is absent, but strong magnetic correlations appear to produce a detectable SSE signal.

We systematically measured the temperature dependence of the SSE for Pt/$BaNi_2V_2O_8$-(100), as shown in Fig. 3(a). In Fig. 3 (b), the slope of the linear fit to the magnetic-field dependence of $V_{2\omega}/P$ is plotted for Pt/$BaNi_2V_2O_8$-(100) and Pt/$BaNi_2V_2O_8$-(001), along with that for Pt/$SiO_2$. Although the SSE for Pt/$BaNi_2V_2O_8$-(100) deviates slightly from the linear magnetic-field dependence at very low temperatures, as shown in Figs. 2(c) and 3(a), we employed a linear approximation in $|\mu_0 H| \leqq 9$ T for all temperatures to ensure a consistent evaluation over the entire temperature range. In Pt/$BaNi_2V_2O_8$-(100), large positive SSE signals are observed at low temperatures, and monotonically decrease as the temperature increases. Interestingly, as is also evident in Fig. 3 (a), it exhibits no significant changes at $T_N$, reminiscent of previous studies of magnetic susceptibility and specific heat in which no anomaly was detected at $T_N$ [35,36].

Upon further increasing the temperature, the SSE signal disappears at approximately 150 K. A significant point is that this temperature is close to $T_{max}$, at which the magnetic susceptibility reaches its maximum. If $T_{max}$ corresponds to the onset temperature of the 2D magnetic character ($J_1 \simeq 12.3$ meV $\simeq 143$ K), the SSE signals can be attributed to strong short-range correlations in $BaNi_2V_2O_8$. In contrast, the SSE signals in Pt/$BaNi_2V_2O_8$-(001) and Pt/$SiO_2$ above $T_N$ are negligibly small; the SSE signals above $T_N$ are unique to Pt/$BaNi_2V_2O_8$-(100).

The long-tailed temperature dependence of the SSE signals even above the transition temperature is considerably different from the SSE reported for typical 3D magnets. For example, for the 3D ferrimagnet YIG, the SSE signals gradually decrease toward the Curie temperature, and disappear above that temperature [44]. For 2D magnets, the SSE was studied for layered ferromagnetic insulators $CrSiTe_3$ and $CrGeTe_3$, and the thermal spin current flowing perpendicular to the 2D layers was measured [32]. The SSE signals were still observed above the Curie temperature and attributed to short-range ferromagnetic correlations reinforced by the Zeeman interaction [32]. However, we point out that the anisotropy in the exchange coupling strength (~ 5 [45]) for these layered magnets is much weaker than that of the present system and the magnetic transition is rather 3D in $CrGeTe_3$ [46]. Moreover, the SSE signals due to the short-range in-plane ferromagnetic correlations for these materials were suppressed by the out-of-plane correlations rapidly diminishing above the Curie temperature, which is in stark contrast to the present case where the SSE disappears almost at the onset temperature of 2D magnetism.

*Discussion*- To understand the observed SSE more deeply, we performed numerical calculations (see also SM [41]). We focus on the tunnel spin current [5,17,18,21–23,47] from $BaNi_2V_2O_8$ to the attached metal Pt to estimate the SSE signals, under the assumption that the interface possesses a weak but finite exchange interaction between localized spins of the magnet and conduction-electron spins of Pt. As the microscopic model for $BaNi_2V_2O_8$ [39], we adopted a 2D Heisenberg antiferromagnet with a small XY anisotropy on a honeycomb lattice. The existence of a field $H$ or anisotropy $D$ reduces the symmetry from SU(2) to U(1) type, and such 2D U(1)-symmetric models exhibit a BKT transition [30,31] like the 2D XY model.

Figure 4(a) shows the numerical result of the magnetic-field dependence of the tunnel spin current at three different temperatures. We find that the spin current monotonically increases with increasing field, and its sign is verified to be the same as that of the ferromagnet (see SM [41]). This is in good agreement with the observed SSE signal in a semi-quantitative level. We also find from Fig. 4(b) that the spin current monotonically decreases with the increase in temperature and even persists in $T > T_{BKT}$, with no characteristic change around $T = T_{BKT}$. This is also quite consistent with the observed SSE signals, and considerably different from the case of 3D ordered magnets. It seems to reflect the weak singularity nature of the BKT transition. These numerical results clearly indicate that the spin current of the SSE in $BaNi_2V_2O_8$ is well described by the spin-wave like excitations in the BKT and paramagnetic phase in the 2D antiferromagnet.

*Conclusion-* We showed that the SSE in a BKT magnet $BaNi_2V_2O_8$ is remarkably different from that in 3D magnets. In Pt/$BaNi_2V_2O_8$-(100), the SSE signals are driven by the spin current flowing within the 2D honeycomb plane of $BaNi_2V_2O_8$, and notably persist well above the Néel temperature, without appreciable change at $T_N$ and $T_{BKT}$. Our numerical calculations indicate that the SSE observed over the entire temperature range is consistent with that expected in a 2D Heisenberg model exhibiting a BKT transition under a magnetic field. Hence, the SSE in the present system plausibly results from strong spin fluctuations, which are characteristic of 2D systems. Although $BaNi_2V_2O_8$ exhibits magnetic order at low temperatures in a static sense, our findings suggest that spin current transport in $BaNi_2V_2O_8$ captures the spin dynamics in BKT-like phase over a very broad temperature range.

*Acknowledgements.* We thank Dr. Takashi Kikkawa and Prof. Hiroto Adachi for the fruitful discussions. This work was carried out by joint research of the Cryogenic Research Center, the University of Tokyo. This work was supported by Japan Science and Technology Agency (JST) FOREST Program, Grant No. JPMJFR203H, and by Japan Society for the Promotion of Science (JSPS) KAKENHI, Grants No. JP22H05449, No. JP22H04464, No. JP23H01832, No. JP23H04576, No. JP23H01392, No. JP21H01794, No. JP23H04582, No. JP20H01830, No. JP20H01849, No. JP19H05825, No. JP19H05600, No. JP22H05131 and No. JP23H04576. K.N. is supported by Research Fellowships of Japan Society for the Promotion of Science for Young Scientists, Grant No. JP22KJ1068. M. K. was supported by JST, the establishment of university fellowships towards the creation of science technology innovation, Grant No. JPMJFS2107.

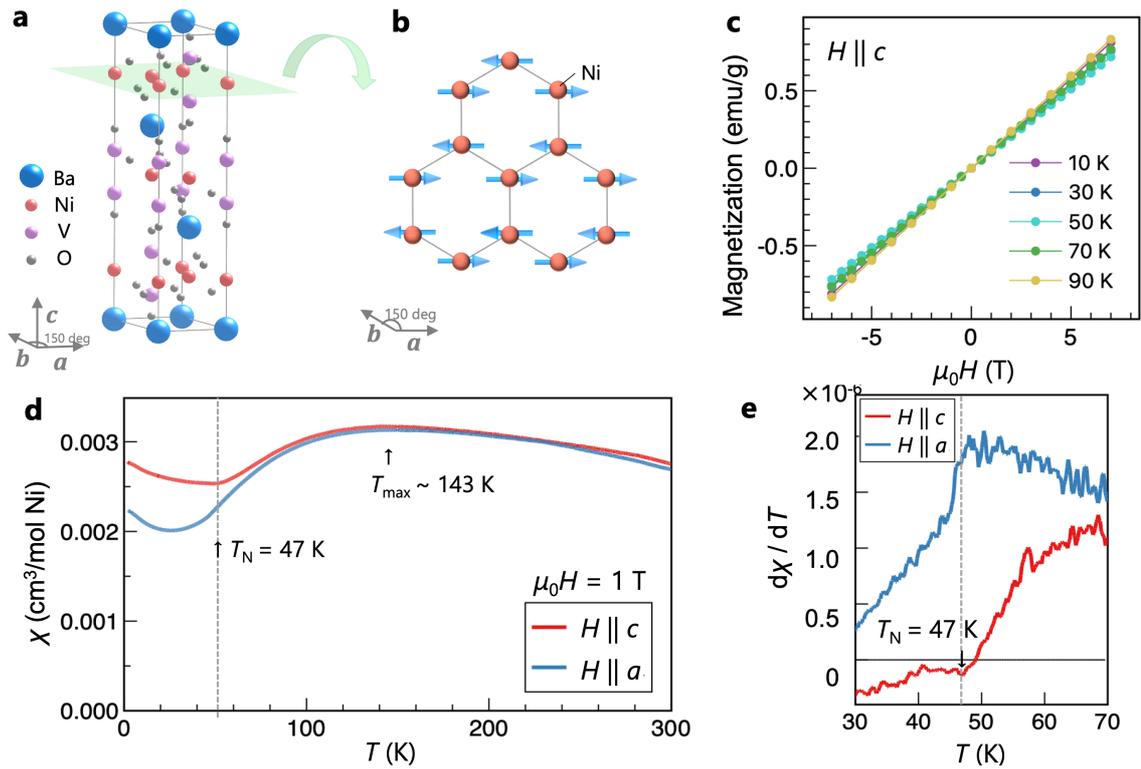

Fig. 1 (a) Crystal structure of BaNi$_2$V$_2$O$_8$. (b) Magnetic structure of a Ni$^{2+}$ honeycomb plane. Each magnetic moment lies approximately in the honeycomb plane. (c) Magnetic-field dependence of magnetization at various temperatures when $H \parallel c$ (i.e., perpendicular to the honeycomb plane). (d), (e) Temperature dependence of (d) magnetic susceptibility $\chi$ and (e) $\frac{d\chi}{dT}$ at 1 T for $H \parallel a$ and $H \parallel c$.

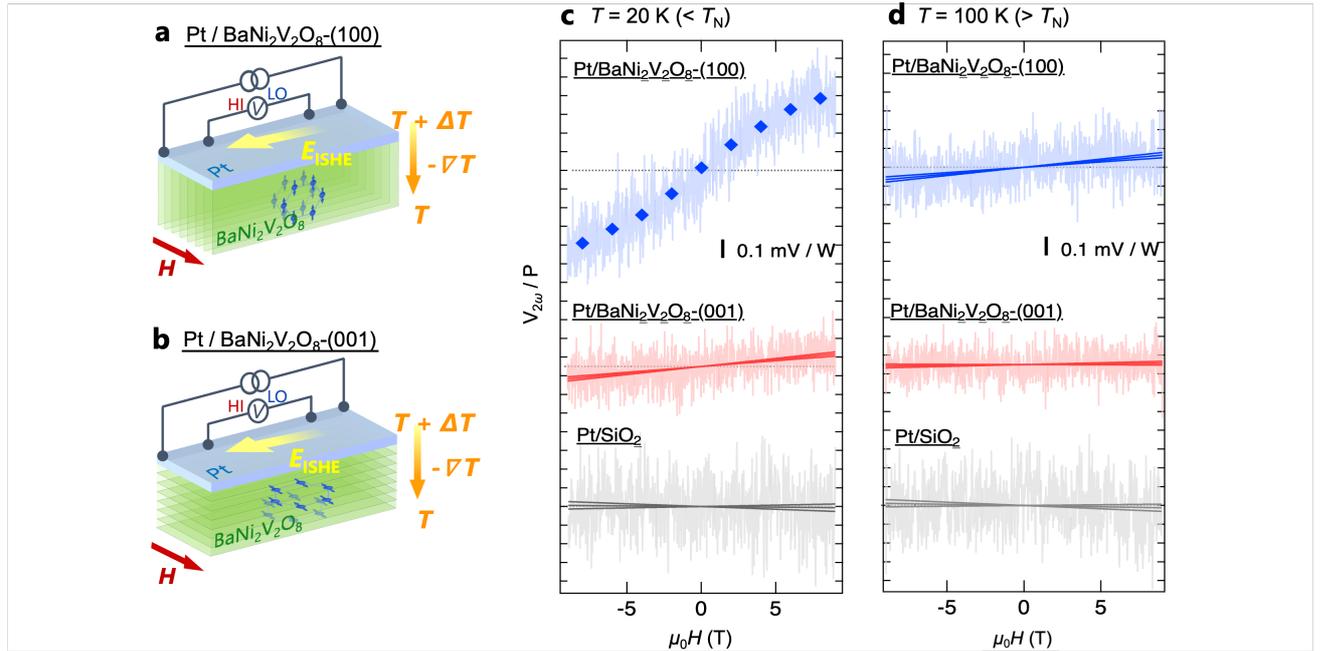

Fig. 2 (a), (b) Schematic illustrations of the SSE measurement setup for (a) Pt/BaNi$_2$V$_2$O$_8$-(100) and (b) Pt/BaNi$_2$V$_2$O$_8$-(001) shown with the 2D honeycomb planes. (c), (d) Magnetic-field dependence of $V_{2\omega}/P$ in Pt/BaNi$_2$V$_2$O$_8$-(100), Pt/BaNi$_2$V$_2$O$_8$-(001), and Pt/SiO$_2$ at (c) 20 K and (d) 100 K. Straight lines represent the linear fits; fitting errors were assessed with the confidence range of 99.73%. The diamond symbols in (c) show the representative points averaged over every 2 T.

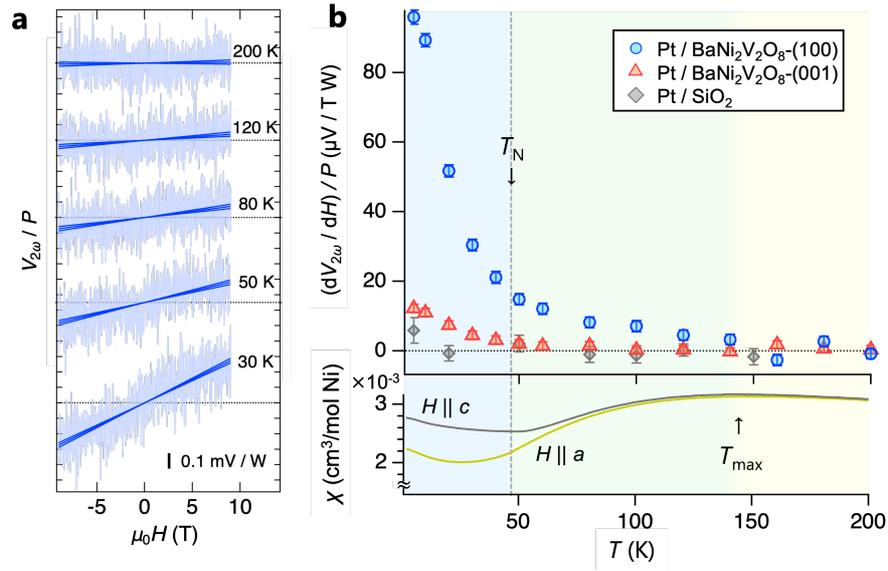

Fig. 3 (a) Magnetic-field dependence of $V_{2\omega}/P$ in Pt/BaNi$_2$V$_2$O$_8$-(100) at various temperatures. (b) Temperature dependence of $(dV_{2\omega}/dH)/P$ in Pt/BaNi$_2$V$_2$O$_8$-(100), Pt/BaNi$_2$V$_2$O$_8$-(001), and Pt/SiO$_2$. The error bars indicate the confidence range of 99.73%. For comparison with the magnetic properties, the temperature dependence of the magnetic susceptibility at 1 T [already shown in Fig. 1 (d)] is redisplayed in the bottom panel.

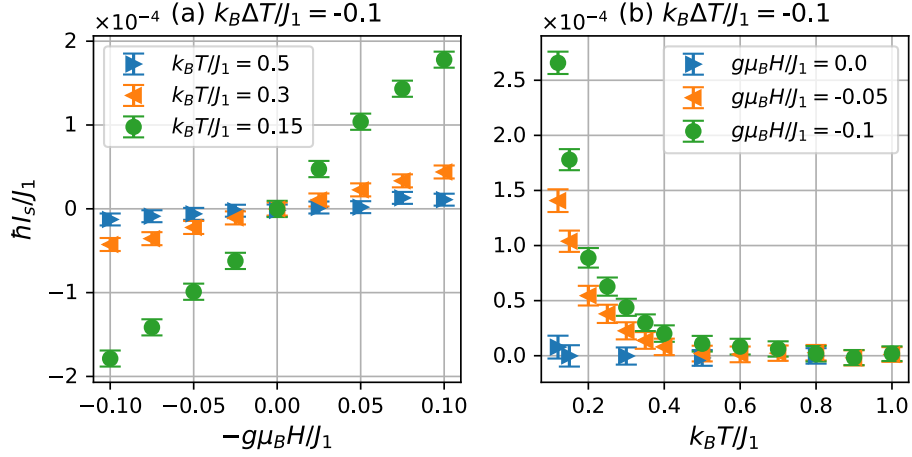

Fig. 4 (a) Magnetic-field $H$ and (b) temperature $T$ dependences of numerically calculated tunnel spin currents in the bilayer model of a 2D antiferromagnet (10 × 10 size) and a metal. Results of larger size systems are obtained in a restricted parameter regime, and they all exhibit the same $H$ and $T$ dependences as the above panels (a) and (b) (see SM [41]). The temperature difference $\Delta T$ is defined as $\Delta T = T_{\text{metal}} - T_{\text{magnet}}$, where $T_{\text{metal}}$ and $T_{\text{magnet}}$ are respectively the temperatures of the metal and the antiferromagnet. Here, $g\mu_B H = \frac{e\hbar}{m}\mu_0 H$. If we set $J_1 \simeq 12.3$ meV $\simeq 143$ K following Ref. [39], $k_B T / J_1 = 0.15$, 0.3, and 0.5 respectively correspond to $k_B T \simeq 21$, 43, and 72 K. In addition, a magnetic field regime $|g\mu_B H| / J_1 < 0.1$ corresponds to $|H| < 10.5$ T. The BKT transition point has been evaluated as $k_B T_{\text{BKT}} \sim 0.3\, J_1$ at $g\mu_B H / J_1 \to 0$ [39]. From previous studies of Refs. [30,31], $k_B T_{\text{BKT}}$ is expected to be around $0.3\, J_1 < k_B T_{\text{BKT}} < 0.6\, J_1$ in a weak-field regime $|g\mu_B H| / J_1 < 0.1$. The error bars indicate the confidence range of 99.73%.

# Supplemental Material: Signature of BKT-like spin transport in a quasi-2D antiferromagnet $BaNi_2V_2O_8$

Kurea Nakagawa,[1] Minoru Kanega,[2] Tomoyuki Yokouchi,[1] Masahiro Sato,[2] and Yuki Shiomi[1]

[1]*Department of Basic Science, The University of Tokyo, Tokyo 153-8902, Japan*

[2]*Department of Physics, Chiba University, Chiba, 263-8522, Japan*

(Dated: December 2023)

### 1. Synthesis of the single crystal of $BaNi_2V_2O_8$

Single crystals of $BaNi_2V_2O_8$ were grown in two steps. First, polycrystalline samples were synthesized using the conventional solid-state reaction. The starting materials $BaCO_3$ (99.99%), NiO (99.97%), and $V_2O_5$ (99.9%) were thoroughly mixed in the stoichiometric ratio and heated in an alumina crucible at 950°C in the air for 2 days. The resultant product was confirmed to be a single phase of $BaNi_2V_2O_8$ by powder x-ray diffraction. Single-crystal samples were then grown by a flux method. Polycrystalline $BaNi_2V_2O_8$ powder was ground and mixed with 5 mol% $V_2O_5$ as a self-flux. The mixture was heated to 1100°C for 10 h, maintained at that temperature for 10 h, slowly cooled to 1000°C at a rate of 1°C/h, and then cooled down to room temperature at a rate of 100°C/h. As described in a previous study [1], single crystals of $BaNi_2V_2O_8$ with yellow hexagonal plates were found and mechanically separated from the flux.

### 2. Source power and frequency dependence of the SSE signals

We show in Fig. S1 the source power [panel (a)] and the source frequency [panel (b)] dependence of the SSE signals in Pt/$BaNi_2V_2O_8$-(100) at 5 K. The SSE magnitude increases almost in proportion to the square of the source current $I$ applied to Pt [Fig. S1 (a)], indicating that the temperature gradient and spin current are driven by Joule heating. The slight deviation from the linear dependence at high currents is attributed to the heating of the entire sample; the increase in the sample temperature reduces the amplitude of the temperature gradient. For the source frequency dependence [Fig. S1 (b)], little voltage dependence on frequency was observed in the measured frequency range of 23 - 83 Hz. As the measurement accuracy of the SSE was slightly better at higher frequencies, all experiments in the main text were conducted at a frequency of 83 Hz.

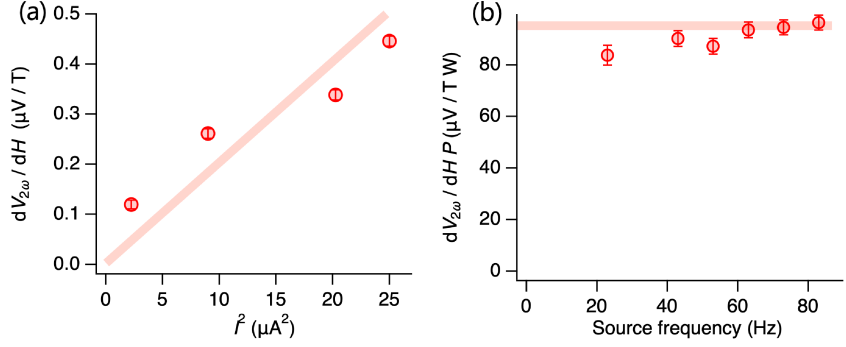

FIG. S1. (a) Source power and (b) source frequency dependence of the SSE in Pt/BaNi$_2$V$_2$O$_8$-(100) at 5 K. The error bars indicate the confidence range of 99.73%.

### 3. Control experiments for Pt/YIG and W/BaNi$_2$V$_2$O$_8$-(100)

To confirm the sign of SSE in Pt/BaNi$_2$V$_2$O$_8$-(100), we conducted a control SSE experiment in Pt/YIG. The SSE signal of Pt/BaNi$_2$V$_2$O$_8$-(100) was confirmed to be the same as that of Pt/YIG [Fig. S2 (a)], a typical ferrimagnetic SSE material. Furthermore, we measured the SSE in W (10 nm)/BaNi$_2$V$_2$O$_8$-(100) to verify the spin current nature of the observed SSE. As shown in Fig. S2 (b), the W (10 nm)/BaNi$_2$V$_2$O$_8$-(100) sample shows the opposite sign of voltage response at 10 K. Because the sign of spin Hall angle of W is opposite to that of Pt [2, 3], the sign reversal of the thermally induced voltages indicates that $V_{2\omega}$ indeed arises from the spin current injected from BaNi$_2$V$_2$O$_8$. In the main text, we focus on Pt/BaNi$_2$V$_2$O$_8$ to study high-temperature SSE signals, because reliable data for W/BaNi$_2$V$_2$O$_8$-(100) were not obtained at high temperatures owing to the relatively high resistance of W.

Model                                                                                                                                3

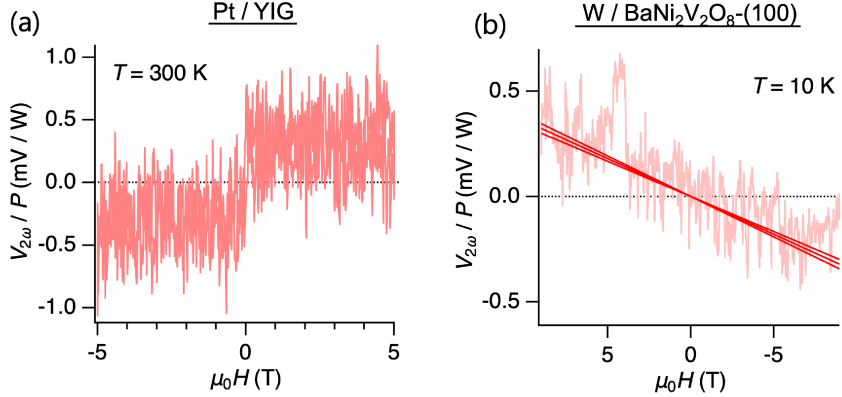

FIG. S2. Magnetic-field dependence of $V_{2\omega}/P$ in (a) Pt/YIG at 300 K and in (b) W/BaNi$_2$V$_2$O$_8$-(100) at 10 K. Straight lines represent the linear fits; fitting errors were assessed with the confidence range of 99.73%.

### 4. Theoretical details of the tunnel spin current

In this section, we explain the details of the formalism of the tunnel spin current [4–6] and its numerical computation.

**Model**

First, we define the model of SSE, a bilayer system consisting of a magnetic insulator and an attached metal. The Hamiltonian of the bilayer system is given by

$$H_{\text{tot}} = H_{\text{magnet}} + H_{\text{metal}} + H_{\text{int}}, \tag{S1}$$

where $H_{\text{magnet}}$ is the Hamiltonian for the magnet, $H_{\text{metal}}$ is that for the metal, and $H_{\text{int}}$ denotes the interfacial interaction between localized spins of the magnet and conducting-electron spins of the metal. As the microscopic model for the magnet BaNi$_2$V$_2$O$_8$ [7], we adopt a 2D Heisenberg antiferromagnet with a small XY anisotropy on a honeycomb lattice, whose Hamiltonian is as follows:

$$H_{\text{magnet}} \equiv H_{\text{2D}} = J_1 \sum_{\langle \bm{r},\bm{r}'\rangle} \bm{S}_{\bm{r}} \cdot \bm{S}_{\bm{r}'} + J_2 \sum_{\langle\langle \bm{r},\bm{r}'\rangle\rangle} \bm{S}_{\bm{r}} \cdot \bm{S}_{\bm{r}'} + D \sum_{\bm{r}} (S_{\bm{r}}^z)^2 - g\mu_{\text{B}} H \sum_{\bm{r}} S_{\bm{r}}^z \tag{S2}$$

where $\bm{S}_{\bm{r}}$ denotes the dimensionless spin-1 operator on site $\bm{r}$ corresponding to each localized spin of Ni$^{2+}$ ion. The first term is the nearest-neighboring exchange interaction with the coupling $J_1$, the second is the next-nearest-neighboring one with the coupling $J_2$, the third is the easy-plane magnetic anisotropy with $D > 0$, and the final term is the Zeeman interaction between spins and external magnetic field $H$, which is in correspondence in the way $g\mu_B H = \frac{e\hbar}{m}\mu_0 H$. The actual crystal structure of BaNi$_2$V$_2$O$_8$ is 3D, but (as we mentioned in the main text) the interlayer coupling is extremely small and we naturally expect that the interlayer coupling is irrelevant to the spin current of SSE. Therefore, the essential features of the SSE can be captured by the purely 2D model $H_{2D}$. The values of coupling constants $J_1$, $J_2$, and $D$ have been estimated [7] and the dominant one is $J_1$: $J_1 \simeq 12.3$ meV, $J_2 \simeq 1.25$ meV, and $D \simeq 0.07$ meV. Other weak interactions have also been considered, but they are negligibly small and here we have ignored them. The existence of a field $H$ or anisotropy $D$ reduces the symmetry from SU(2) to U(1) type, and such 2D U(1)-symmetric models exhibit a BKT transition [8–10] like the 2D XY model. In the model of $H_{2D}$ with $g\mu_B H/J_1 \to 0$, the BKT transition temperature has been estimated as $k_B T_{BKT} \simeq 45$ K [7] for $J_1 \simeq 12.3$ meV $\simeq 143$ K.

The interfacial interaction is assumed to be a simple exchange type:

$$H_{\text{int}} = \sum_{\boldsymbol{r}_{\text{int}}} J_{\boldsymbol{r}_{\text{int}}} \boldsymbol{S}_{\boldsymbol{r}_{\text{int}}} \cdot \boldsymbol{T}_{\boldsymbol{r}_{\text{int}}}, \tag{S3}$$

where $\boldsymbol{S}_{\boldsymbol{r}_{\text{int}}}$ and $\boldsymbol{T}_{\boldsymbol{r}_{\text{int}}}$ are respectively a spin of the magnet and a conducting-electron spin of the metal on an interfacial site $\boldsymbol{r}_{\text{int}}$. The strength of the coupling constant $J_{\boldsymbol{r}_{\text{int}}}$ is assumed to be randomly distributed because interfacial crystal structure is usually misfit.

If one accurately describes the spin dynamics of the metal, the Hamiltonian $H_{\text{metal}}$ should be chosen to a free (or weakly interacting) electron model. However, it is generally hard (even by using numerical methods) to compute the nonequilibrium time evolution of the coupled system $H_{\text{tot}}$ in fully quantum, microscopic level at finite temperatures. Moreover, the present study focuses on the 2D characteristic features of the magnet BaNi$_2$V$_2$O$_8$, and does not on the detailed properties of the metal. From these arguments, we adopt a set of independent single-site models to extract only the important nature of the metal. Within this model, conducting electrons are not directly correlated with each other and the Hamiltonian $H_{\text{metal}} = 0$. The model for the metal will be rigidly defined in the form of the equation of motion (EOM) soon later. Our modelling of $H_{\text{tot}}$ is summarized in Fig. S3.

### Tunnel spin current

Next, we discuss the tunnel spin current injected from the magnet to the metal. This current is generated by a temperature gradient perpendicular to the interface. In a real setup, the temperature gradually changes along the direction from the magnet to the metal. However, for simplicity, we discretize the temperature distribution: The magnet has temperature $T_{\rm magnet} = T$ and the metal is in $T_{\rm metal} = T + \Delta T$. The difference $\Delta T = T_{\rm metal} - T_{\rm magnet}$ is the driving force of the spin current. A magnetic field $H$ along the $z$ direction makes the polarization of the spin current parallel to the $z$ direction. Under these preparations, the tunnel spin current can be defined by the time derivative of the $z$ component of total spin in the metal. Therefore, the spin current is given by

$$I_s = \sum_{\bm{r}} \partial_t \langle T^z_{\bm{r}} \rangle = \sum_{\bm{r}_{\rm int}} \frac{1}{\hbar} \langle J_{\bm{r}_{\rm int}} (T^x_{\bm{r}_{\rm int}} S^y_{\bm{r}_{\rm int}} - T^y_{\bm{r}_{\rm int}} S^x_{\bm{r}_{\rm int}}) \rangle, \tag{S4}$$

where $\langle \cdots \rangle$ denotes the statistical average for the nonequilibrium steady state of $H_{\rm tot}$, which includes the average about the random distribution of $J_{\bm{r}_{\rm int}}$. In the second equality, we have used the Heisenberg's EOM for $T^z_{\bm{r}}$ and the interfacial interaction leads to the product between $T^\alpha_{\bm{r}_{\rm int}}$ and $S^\beta_{\bm{r}_{\rm int}}$. The tunnel spin current $I_s$ is proportional to the SSE signal (i.e., electric voltage), and therefore, we can theoretically predict the field and temperature dependences of the SSE signal from the spin current. This formalism can be applicable to a broad class of magnetic systems and in fact has succeeded in explaining several SSE signals in different magnets such as a 1D spin liquid [11], a spin-nematic liquid [12], and a spin-Peierls magnet [13].

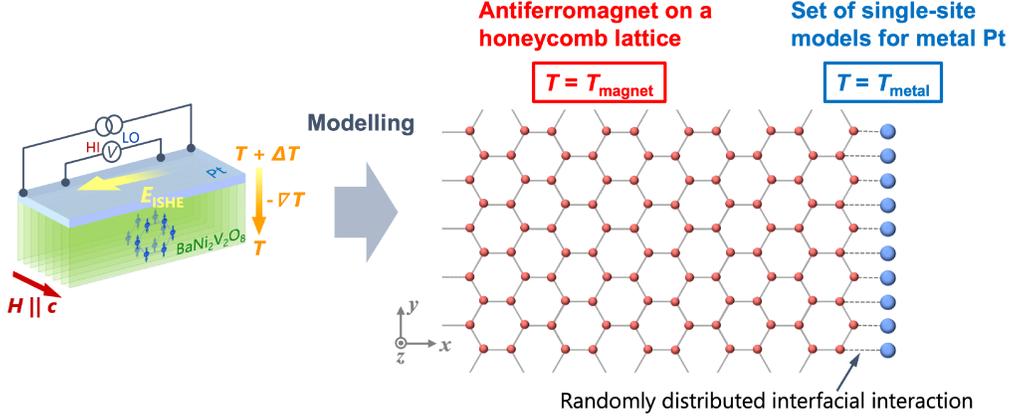

FIG. S3. Schematics of our modelling. The left panel illustrates the experimental setup of the SSE in the present study and the right one shows our model consisting of an antiferromagnetic Heisenberg model of a 2D honeycomb lattice (the model for BaNi$_2$V$_2$O$_8$) and a set of single-site (red circles) systems (the model for the metal). Each site of the metal is located at a boundary of the honeycomb lattice. The interfacial coupling constant $J_{\bm{r}_{\rm int}}$ on each site $\bm{r}_{\rm int}$ is chosen to be a random variable.

The remaining task is to compute the right-hand side of Eq. (S4). In the present work, we concentrate on the spin dynamics at finite temperatures, especially, around the BKT transition at $T = T_{\rm BKT}$. Therefore, the thermal fluctuation effect on spins is more important than the quantum fluctuation. In addition to it, we should sufficiently take into account the thermal fluctuation effects in both spatial and temporal directions to correctly describe the spin dynamics around $T = T_{\rm BKT}$. Namely, we have to go beyond mean-field type methods such as spin-wave [4, 6] and Ginzburg-Landau [14] theories. We thus compute the spin dynamics of the magnet $H_{\rm 2D}$ by numerically solving the stochastic Landau-Lifshitz-Gilbert (LLG) equation [15] at finite temperatures. The LLG equation fully captures such thermal fluctuation effects. The equation is given by

$$\hbar \frac{d}{dt} \bm{S_r} = -\bm{S_r} \times \left( \frac{\partial H_{\rm tot}}{\partial \bm{S_r}} + \bm{h_r}(t) \right) - \alpha_{\rm magnet} \frac{\bm{S_r}}{S} \times \hbar \frac{d}{dt} \bm{S_r}, \tag{S5}$$

where the first term represents the spin dynamics driven by the torque (effective magnetic field) $\frac{\partial H_{\rm tot}}{\partial \bm{S_r}} = (\frac{\partial H_{\rm tot}}{\partial S_r^x}, \frac{\partial H_{\rm tot}}{\partial S_r^y}, \frac{\partial H_{\rm tot}}{\partial S_r^z})$ and a random field $\bm{h_r}(t)$, and the second term is a Gilbert damping with a dimensionless constant $\alpha_{\rm magnet}$. In magnetic materials, $\alpha_{\rm magnet} \sim 10^{-2} - 10^{-5}$. We note that in this stage, $\bm{S_r}$ stands for a three-component vector with length $S = 1$ (it is no longer the spin operator). The random field is chosen to be a Gaussian white noise satisfying

$$\langle h_{\bm{r}}^\alpha(t) \rangle = 0, \tag{S6}$$

$$\langle h_{\bm{r}}^\alpha(t) h_{\bm{r}'}^\beta(t') \rangle = 2\hbar \alpha_{\rm magnet} k_{\rm B} T_{\rm magnet} \delta_{\alpha,\beta} \delta_{\bm{r},\bm{r}'} \delta(t-t'). \tag{S7}$$

The random field and the Gilbert damping term cooperatively make the system relax to the equilibrium state with $T = T_{\rm magnet}$.

Considering Eq. (S5), let us define the EOM for the electron spins of the metal. We here make conducting-electron spins obey the following LLG equation

$$\hbar \frac{d}{dt} \bm{T}_{\bm{r}_{\rm int}} = -\bm{T}_{\bm{r}_{\rm int}} \times \left( \frac{\partial H_{\rm tot}}{\partial \bm{T}_{\bm{r}_{\rm int}}} + \bm{g}_{\bm{r}_{\rm int}}(t) \right) - \alpha_{\rm metal} \frac{\bm{T}_{\bm{r}_{\rm int}}}{\mathcal{T}} \times \hbar \frac{d}{dt} \bm{T}_{\bm{r}_{\rm int}}, \tag{S8}$$

where the three-component vector $\bm{T}_{\bm{r}_{\rm int}}$ with length $\mathcal{T}$ represents the conducting-electron spin on an interfacial site $\bm{r}_{\rm int}$, $H_{\rm metal} = 0$ in the total Hamiltonian $H_{\rm tot}$, $\bm{g}_{\bm{r}_{\rm int}}(t)$ is the random field, and $\alpha_{\rm metal}$ is the dimensionless Gilbert damping constant for the metal. In order to make the metal relax to its equilibrium state with $T = T_{\rm metal}$, we assume that the random force obeys

$$\langle g_{\bm{r}}^\alpha(t) \rangle = 0, \tag{S9}$$

$$\langle g_{\bm{r}}^\alpha(t) g_{\bm{r}'}^\beta(t') \rangle = 2\hbar \alpha_{\rm metal} k_{\rm B} T_{\rm metal} \delta_{\alpha,\beta} \delta_{\bm{r},\bm{r}'} \delta(t-t'). \tag{S10}$$

In general, the spin relaxation time of metals is much shorter than that of magnets. Hence we make $\alpha_{\rm metal}$ larger than $\alpha_{\rm magnet}$. The LLG equation for a spin-$S$ magnet $H_{\rm magnet}$ can be exactly mapped to another LLG equation for a spin-$S'$ magnet ($S' \neq S$) with the same form of the Hamiltonian if we perform a proper re-normalization of time $t$, field $H$, and temperature $T$. However, this nature is violated for a bilayer system when $\bm{S_r}$ and $\bm{T}_{\bm{r}_{\text{int}}}$ have different spin magnitudes ($S \neq \mathcal{T}$). To avoid the violation, we simply set $\mathcal{T} = S$.

Finally, we comment on one technical aspect regarding the calculation of the spin current. In previous studies based on the formula of the tunnel spin current, the researchers have often approximated Eq. (S4) by applying the perturbation theory with respect to a weak interfacial interaction $H_{\text{int}}$ [4, 6, 11–13]. As a result, the spin current is given by

$$I_s \propto \Delta T J_{\bm{r}_{\text{int}}}^2 \int_{-\infty}^{\infty} d\omega \, \text{Im}[\chi_{\text{magnet}}^R(\omega)] \, \text{Im}[\chi_{\text{metal}}^R(\omega)] \, \frac{1}{(k_{\text{B}}T)^2} \frac{(\hbar\omega)^2}{\sinh^2(\frac{\hbar\omega}{2k_{\text{B}}T})}, \quad (S11)$$

up to the first order perturbation with respect to $J_{\bm{r}_{\text{int}}}$. Here, $\chi_{\text{magnet}}^R(\omega)$ and $\chi_{\text{metal}}^R(\omega)$ are respectively the retarded parts of dynamical susceptibilities for the magnet at $T_{\text{magnet}} = T$ and the metal at $T_{\text{metal}} = T + \Delta T$, $T = (T_{\text{magnet}} + T_{\text{metal}})/2$ (for $\Delta T \to 0$) is the mean value of temperature, and $\omega$ is the angular frequency. These susceptibilities can be evaluated by using techniques of equilibrium statistical mechanics. However, this perturbation theory is not applicable in the present study. We have replaced spin operators with classical vectors to fully involve the thermal fluctuation effect. In such classical spin models, all retarded Green's functions vanish since all the commutators in the Green's functions are zero. Therefore, instead of the formula of Eq. (S11), we have utilized a numerical non-perturbative method to estimate the spin current $I_s$.

## Numerical calculation: Field and temperature dependences

Here, we explain some details and results of the numerical computation of Eq. (S4). We consider finite-size 2D models $H_{\text{2D}}$ on $L \times L$ sites honeycomb lattice. The linear length $L$ is chosen to be 10, 20, 40, and 80. We introduce $L$ conducting-electron spins $\bm{T}_{\bm{r}_{\text{int}}}$ for a $L \times L$ magnet and align the positions of $\bm{r}_{\text{int}}$ along an 1D boundary of the $L \times L$ honeycomb lattice (see Fig. S3). The magnitude of $J_{\bm{r}_{\text{int}}}$ is uniformly distributed in the range of $0 < J_{\bm{r}_{\text{int}}} < 0.3 J_1$. The average $\langle \cdots \rangle$ is taken by two steps: The ensemble average is taken by using $\mathcal{O}(10^{6-7})$ ensembles, and the time average is computed in the range $0 < J_1 t/\hbar < 100$ for each ensemble. Note that we first wait for a long enough time so that the bilayer system approaches the nonequilibrium steady state, and then we take the time average. It is difficult to perform the numerical calculation of $I_s$ in large systems ($L \gtrsim 50$) for a broad parameter space of $(H, k_{\text{B}}T)$ from the aspect of the numerical cost. However, as we will discuss in the next subsection, we observe that even in the $L = 10$ system, the field and temperature dependences of $I_s$ exhibit the same tendency as those in larger systems with $L = 40$ and 80. Therefore, we mainly show the result of $L = 10$ in a broad parameter regime.

Figure S4 (a) and (b) respectively show the magnetic-field and temperature dependences of the tunnel spin current. First, we can verify that $I_s$ vanishes in the case of $H = 0$ or $\Delta T = 0$. Secondly, we find that $I_s$ changes its sign when $H$ (or $\Delta T$) is changed into $-H$ ($-\Delta T$). These behaviors should be satisfied from the viewpoint of thermodynamics. We note that the asymmetric relation $I_s(H) = -I_s(-H)$ is somewhat broken in panel (a). This is because our numerical calculation is done under the condition of a finite $\Delta T$, and temperature has the lower bound $T = 0$. If we sufficiently approach to the limit $\Delta T \to 0$, the asymmetry is expected to be recovered. As we discussed in the main text, the magnetic-field and temperature dependences are in good agreement with the experimental results of $BaNi_2V_2O_8$.

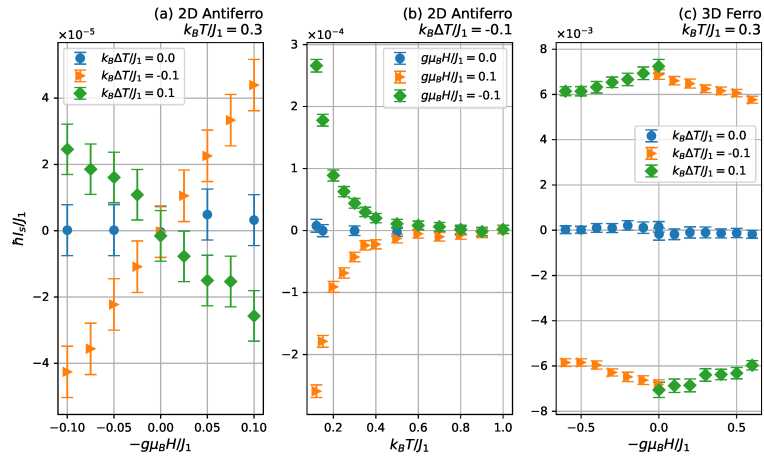

FIG. S4. (a) Magnetic-field $H$ and (b) temperature $T$ dependences of the numerically calculated tunnel spin current in the bilayer model of a 2D antiferromagnet ($L = 10$) and metal. (c) Field dependence of the tunnel spin current in a bilayer model of 3D $S = 1$ ferromagnet $H_{3D}$ ($10 \times 10 \times 5$) and a metal. The error bars indicate the confidence range of 99.73%. In the case of $H_{2D}$ [panels (a) and (b)], we set $J_1 = 1$, $J_2 = 0.102$ and $D = 0.00565$, following Ref. [7]. The Gilbert damping constants are chosen to be $\alpha_{\text{magnet}} = 0.05$ and $\alpha_{\text{metal}} = 0.5$. If we set $J_1 \simeq 12.3$ meV $\simeq 143$ K, $k_B T/J_1 = 0.15$, 0.3, and 0.5 respectively correspond to $k_B T \simeq 21$, 43, and 72 K. In addition, the magnetic field regime $|g\mu_B H|/J_1 < 0.1$ corresponds to $|H| \lesssim 10$ T. The BKT transition point has been evaluated as $k_B T_{\text{BKT}} \sim 0.3 J_1$ at $g\mu_B H/J_1 \to 0$ [7]. From previous studies of Refs. [8, 9], $k_B T_{\text{BKT}}$ is expected to be around $0.3 J_1 < k_B T_{\text{BKT}} < 0.6 J_1$ in a weak-field regime $|g\mu_B H|/J_1 < 0.1$. In the case of $H_{3D}$ [panel (c)], we set $J_1 = 1$ and $D_z = 0.1$. The damping constants are the same as those of the 2D case. The model of the metal consists of $10 \times 10$ sites and is coupled to the 3D magnet through the surface of the magnet.

As a comparison to the 2D model $H_{2D}$, we also compute the tunnel spin current in another bilayer system of a 3D ferromagnetic Heisenberg model $H_{3\mathrm{D}}$ on a cubic lattice and a metal. The Hamiltonian $H_{3\mathrm{D}}$ is given by

$$H_{3\mathrm{D}} = -J_1 \sum_{\langle \bm{r},\bm{r}' \rangle} \bm{S}_{\bm{r}} \cdot \bm{S}_{\bm{r}'} - D_z \sum_{\bm{r}} (S_{\bm{r}}^z)^2 - g\mu_{\mathrm{B}} H \sum_{\bm{r}} S_{\bm{r}}^z \qquad (\mathrm{S}12)$$

where the first term represents the nearest-neighboring ferromagnetic exchange interaction with a coupling constant $J_1 > 0$, the second is an easy-axis anisotropy with $D_z > 0$, and the final term is the Zeeman interaction by an external magnetic field $H$ along the $z$ direction. We define the total Hamiltonian by replacing $H_{2\mathrm{D}}$ with $H_{3\mathrm{D}}$ in $H_{\mathrm{tot}}$. The model for the metal is given by a set of single-site models, similar to that of the previous bilayer system. However, we note that in the 3D model, each site of the metal is coupled to the cubic lattice 3D magnet on a certain 2D surface (not along the 1D line). The 3D ferromagnetic ordered state is stable and the thermal fluctuation effect is much weaker than that in the 2D antiferromagnet if the temperature is sufficiently low in the ordered phase. Therefore, a small ensemble number and a small linear size $L$ are enough to take the average $\langle \cdots \rangle$ in the 3D ordered case. Figure S4 (c) represents the result of $I_s$ in the 3D ferromagnet. The field and temperature dependences are completely different from those of the 2D antiferromagnet. And they are very consistent with the well-known experimental SSE signals of 3D ferromagnets [16, 17]. This result of $H_{3\mathrm{D}}$ indicates that our modelling well works to describe the tunnel spin currents in SSE setups.

### Numerical calculation: System-size dependence

In this subsection, we discuss the finite-size effect of our numerical calculations. We focus on the bilayer system of the 2D model $H_{2\mathrm{D}}$, in which the fluctuation effect is very important. Figure S5 (a) shows the field dependence of the spin current $I_s$ in the bilayer systems of different sizes $L$. The values of $L = \infty$ are obtained by extrapolating $I_s$ in the thermodynamic limit from results of $L = 10, 20, 40,$ and $80$. The extrapolation process is depicted in Fig. S5 (b). From these results, we find that even the spin current in the $L = 10$ system exhibits the same tendency as those in larger systems.

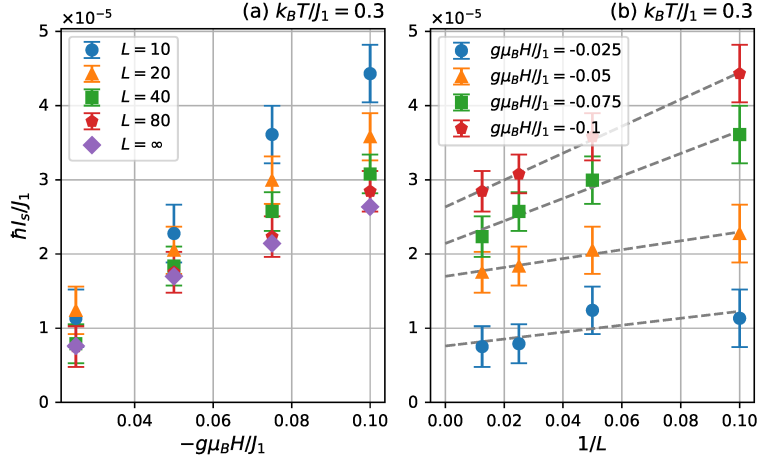

FIG. S5. System-size dependence of the computed tunnel spin current in the bilayer systems of $H_{2D}$ and the metal. Panel (a) represents the magnetic-field dependence of the spin current $I_s$ in different-size systems. Panel (b) shows the extrapolation method to obtain the result in the thermodynamic limit $L = \infty$. The dotted lines are determined by fitting numerical results with a function $f(L) = c_0 + c_1/L$. The error bars show the confidence range of 99.73%.

### Order parameters

In this final subsection, we verify that there is no magnetic long-range order in our numerical simulation for $H_{2D}$. Figures S6 (a) and S6 (b) show the typical time evolution of the ensemble averages of uniform magnetization $M^\alpha = (S^\alpha_{\bm{r}_A} + S^\alpha_{\bm{r}_B})/2$ and staggered one $N^\alpha = (S^\alpha_{\bm{r}_A} - S^\alpha_{\bm{r}_B})/2$ in a certain unit cell including an A-sublattice site $\bm{r}_A$ and a B-sublattice one $\bm{r}_B$. The figures also show the standard deviation of the magnetizations, $\sigma(M^\alpha) = (\langle (M^\alpha)^2 \rangle - \langle M^\alpha \rangle^2)^{1/2}$ and $\sigma(N^\alpha) = (\langle (N^\alpha)^2 \rangle - \langle N^\alpha \rangle^2)^{1/2}$, in which the symbol $\langle \cdots \rangle$ means the ensemble average. We start from the fully polarized state with $S^z_{\bm{r}} = +1$ in the time evolution. One finds that the ensemble averages of both $M^{x,y}$ and $N^{x,y,z}$ almost vanish at both low and moderate temperatures, and the computed $\sigma(N^\alpha)$ show that the in-plane Néel order parameters $N^{x,y}$ strongly fluctuate around their mean values at a low temperature. These results indicate that our numerical solution of the LLG equation successfully simulates the quasi-long-range ordered BKT phase of $H_{2D}$ in the low-temperature regime.

As a comparison, we also show time evolution of the ensemble-averaged magnetization $\langle M^\alpha \rangle = \langle S^\alpha_{\bm{r}} \rangle$ on a single site $\bm{r}$ in the 3D cubic-lattice ferromagnet $H_{3D}$ in Fig. S6 (c). Comparing Fig. S6 (a) and S6 (c), we find that the 2D system $H_{2D}$ exhibits a much larger fluctuation of magnetic

Order parameters 11orders rather than those of the 3D ordered state of $H_{3D}$. Figure S6 (c) also shows that the 3D ferromagnetic long-range order survives in the long-time evolution at a sufficiently low temperature.

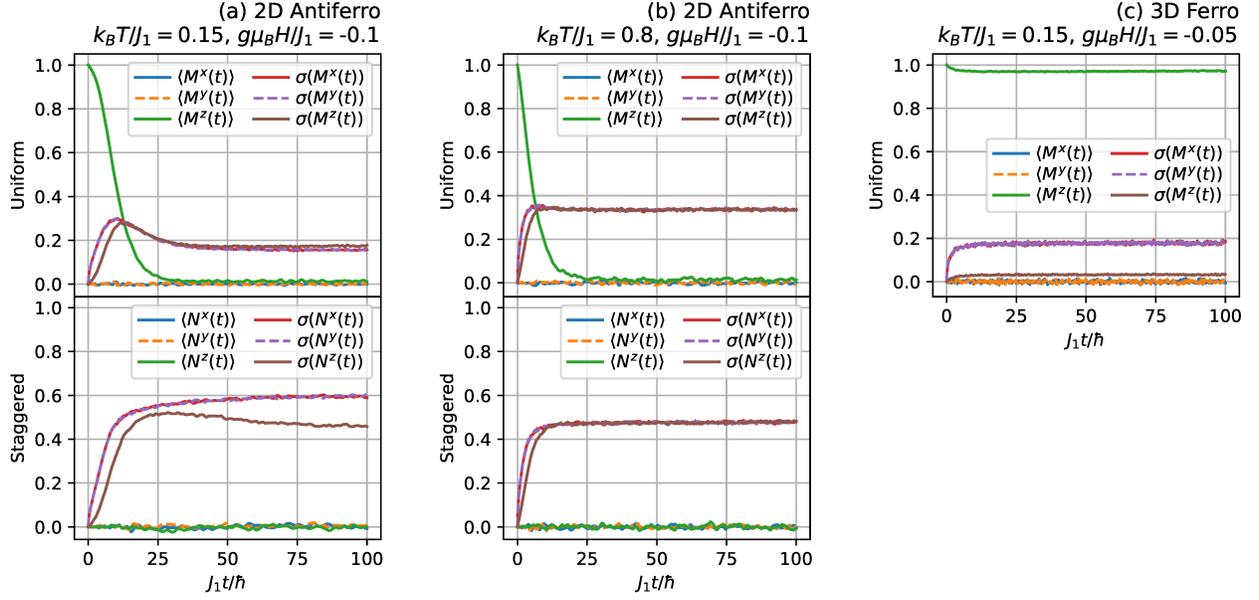

FIG. S6. Time evolution of averaged values of uniform and staggered magnetizations on a single unit cell, $\langle M^\alpha \rangle$ and $\langle N^\alpha \rangle$, in the 2D antiferromagnet $H_{2D}$ at (a) low and (b) moderate temperatures. The panels (a) and (b) also plot their standard deviations $\sigma(M^\alpha)$ and $\sigma(N^\alpha)$. The system size is chosen to be $80 \times 80$, and we have used 5000 ensembles to take the average. As a comparison, we also depict (c) the averaged single-site magnetization $\langle M^\alpha \rangle$ and its standard deviation $\sigma(M^\alpha)$ for the 3D ferromagnet $H_{3D}$. The system size is $10 \times 10 \times 10$ and we have prepared 1000 ensembles for averaging. We note that in all the numerical simulations, we do not attach the model of the metal, i.e., we calculate the time evolution of localized spins on the purely magnetic models $H_{2D}$ and $H_{3D}$ in these figures.